\documentclass[graybox]{svmult}

\usepackage{mathptmx}        % selects Times Roman as basic font
\usepackage{helvet}          % selects Helvetica as sans-serif font
\usepackage{courier}         % selects Courier as typewriter font
\usepackage{type1cm}         % activate if the above 3 fonts are
\usepackage{makeidx}         % allows index generation
\usepackage{graphicx}        % standard LaTeX graphics tool
\usepackage{multicol}        % used for the two-column index
\usepackage[bottom]{footmisc}% places footnotes at page bottom
\usepackage{amsmath,amsfonts,amssymb}
\usepackage[all]{xy}
\usepackage{listings,color}
\usepackage{url}
\usepackage[noadjust]{cite}
\definecolor{backgroundColor}{rgb}{0.85,0.85,1}
\lstdefinelanguage{maple}{%
  morekeywords={ dummy, proc, begin, end, if, fi, then, else, do,%
                cmul, wedge, eval, subs, cmulGTensor, cmulTensor,%
                switch, gswitch, local, phi, evalm, asvd, cm, ID,%
                Id, CM, error, global, return, displayid, type,%
                dummy}
  morecomment=[l]{\#}
  }
\makeindex                   % used for the subject index
\def\Clifford{\texttt{CLIFFORD}}
\def\Bigebra{\texttt{Bigebra}}
\def\Maple{\texttt{Maple}}
\def\CL{C\!\ell}
\def\Mat{\mathrm{Mat}}
\def\rev{\mathrm{rev}}
\def\End{\mathrm{End}}
\def\code#1{\texttt{\small #1}}
\DeclareMathOperator{\hotimes}{\hat{\otimes}\,}
\newcommand{\ed}{\end{document}}
\newcommand{\rcite}[1]{{\rm \cite{#1}}}
\newcommand{\spn}{{\rm span}}
\newcommand{\JJ}{\mathbin{\raisebox{0.10ex}{$\footnotesize
                       \rm\vphantom{I}%
                       \_\hskip -0.25em\_%
                       \vrule width 0.6pt$}}\,}           %left contraction
\begin{document}
\title*{Using Periodicity Theorems for Computations in Higher
Dimensional Clifford Algebras}
\titlerunning{Symbolic Computations in Higher Dimensional Clifford Algebras}
\author{Rafa{\l} Ab{\l}amowicz and Bertfried Fauser}
\institute{Rafa{\l} Ab{\l}amowicz \at 
Department of Mathematics, Box 5054,
Tennessee Technological University,
Cookeville, TN 38505, U.S.A. 
\email{rablamowicz@math.tntech.edu}
\and
Bertfried Fauser \at 
The University of Birmingham, 
School of Computer Science,
Edgbaston-Birmingham, B15 2TT, 
England.
\email{b.fauser@cs.bham.ac.uk}}
\maketitle
%
%\abstract*{--same abstract as below??--}
%%% vskip trickery here 
%\vskip-0in
\abstract{We present different methods for symbolic computer algebra
computations in higher dimensional ($\ge9$) Clifford algebras using
the \Clifford\ and \Bigebra\ packages for \Maple\textregistered.
This is achieved using graded tensor decompositions, periodicity
theorems and matrix spinor representations over Clifford numbers. We
show how to code the graded algebra isomorphisms and the main
involutions, and we provide some benchmarks.}

\section{Introduction}\label{sec:introduction}
Clifford algebras are used in several areas of mathematics, physics,
and engineering. Since computing power has increased tremendously, in
terms of available memory and actual processing speed, practical
symbolic computations, say on a laptop finishing within minutes, are
by now possible for Clifford algebras over vectors spaces of dimensions
higher than~$8$. When \Clifford\ was designed, way back in the early
90's~\cite{A:1996}, such computations where impossible. However, the
design of \Clifford, restricting the vector space dimension to less
than or equal to $9$, incorporated from the beginning the idea that
using mod-$8$ and other periodicity isomorphisms of real Clifford
algebras allows nevertheless computations in higher dimensional
algebras as well. Using periodicity theorems also `groups' Clifford
algebras as in a periodic table, the \emph{spinorial chess
board}~\cite{budinich:trautmann:1988a}. Hence, implementing  the
periodicity in symbolic computations --as described in this paper--
makes use of these intrinsic algebra features. Recent applications in
engineering, for example when modeling geometric transformations in
robotics, rely on real Clifford algebras like
$\CL_{8,2}$~\cite{bayro-corrochano:2012} and thus enforce the need for
using the periodicity in computations with \Clifford.

In this note, we will describe three ways showing how \Clifford\ deals
with (real) Clifford algebras of higher dimensions ($\ge9$). One way is
to use \Bigebra, an extension of \Clifford, to utilize graded tensor
products of Clifford algebras. Another method utilizes ungraded tensor
products and periodicity isomorphisms without a need for introducing
spinorial bases. The third method uses, on top of the periodicity
theorems, a spinor representation of one factor of the tensor
decomposition, hence computing in Clifford algebra valued matrix rings. For
example, from $\CL_{p+1,q+1}\simeq \CL_{p,q}\otimes \CL_{1,1}
\simeq \Mat(2,\CL_{p,q})$ we see that computing say in $\CL_{8,2}$ can
be done with the $2 \times 2$ matrices over $\CL_{7,1}$. \Clifford\
supports computing with matrices having Clifford entries. Similar
computations still make sense, e.g., for conformal symmetries using
Vahlen matrices $\Mat(2,\CL_{3,1})\simeq \CL_{4,2}$.

After recalling our basic notations and the periodicity theorems, we
start explaining how we use \Bigebra\ to compute with graded and
ungraded tensor products of Clifford algebras. Then we proceed to the
third  matrix-based method indicated above. As this method relies on
spinor representations of real Clifford algebras, one encounters not
only real, but also complex, quaternionic, double real, and double
quaternionic spinor representations. For the sake of simplicity and
space, we will just deal with real representations of simple Clifford
algebras. The periodicity theorems single out the signature cases
where a graded algebra isomorphism is available onto an ungraded 
tensor decomposition and hence, introducing spinor bases, matrix tensor
products can be employed.

\section{Tensor product decompositions and periodicity for $\CL$-algebras}
\label{sec:1}
We study Clifford algebras over finite (real) vector spaces. While these
algebras can be defined by a universal property, for actual computations
in a CAS (Computer Algebra System) we use generators and relations.

\subsection{Basic notations, quadratic and bilinear forms}
\label{subsec:notations}
Let $V$ be a finite dimensional real (or complex) vector space with
scalar multiplication $\mathbb{R} \times V \rightarrow V :: 
(\lambda,v) \mapsto \lambda v$. A \emph{quadratic form} is a map 
$Q : V \rightarrow \mathbb{R}$ such that $Q(\lambda v) = \lambda^2 Q(v)$,
with associated polar bilinear form $B : V\times V \rightarrow \mathbb{R}$
derived from $Q$ and defined as $2B(v,w) = Q(v+w) - Q(v) - Q(w)$ (hence
$Q(v)=B(v,v)$). $Q$ is called \emph{non-degenerate} if $Q(v)=0$ implies
$v=0$ ($v\in V, \forall w\in V, B(w,v)=0$ implies $v=0$). An isomorphism
$V\simeq \mathbb{R}^n=\oplus^n\mathbb{R}$ defines a set of generators
for $V$ from the injections of the direct sum
$i_k : \mathbb{R}\rightarrow \mathbb{R}^n :: 1\mapsto e_k\simeq v_k$,
that is a basis for $V$. Sylvester's theorem states that there exists
a basis for $V$ such that the quadratic form~$Q$ is diagonal with
entries $\pm 1,0$ in the real case ($0$ only when $Q$ is degenerate; just $+1$'s
in the non-degenerate complex case). Under these isomorphisms, the real
quadratic space $(Q,V)$ is isomorphic to a space
$\mathbb{R}^{p,q}=(\mathbb{R}^n,\delta_{p,q})$ with a diagonal quadratic
(polarized) form  $\delta_{p,q}$ with $p$ ones and $q$ minus ones.
$(p,q)$ is called the \emph{signature} of $Q$ and $p+q=n = \dim V$.
Furthermore we have generators $e_k$ of $\mathbb{R}^{p,q}$ such that
$Q(e_k)=+1$ for $1\le k\le p$ and $Q(e_k)=-1$ for $p<k\le p+q$.

\subsection{Grassmann algebra, $\mathbb{Z}$- and $\mathbb{Z}_2$-gradings, main involutions}
\label{susec:grassmann}
We can associate functorially to any (say finite, real) vector space $V$
the Grassmann algebra of antisymmetric tensors,
$\bigwedge : V\rightarrow \bigwedge V = \oplus_{i=0}^n \bigwedge^i V$.
Using the linearly ordered basis $\{e_k\}$ of $V$ ($e_k <e_l$ if $k<l$),
we get a basis 
$$
  \{e_I\} 
   =\{ e_{i_1}\wedge e_{i_2} \wedge \cdots \wedge e_{i_l} \mid  I
   =(i_1,i_2,\ldots,i_l),\, i_1<i_2<\cdots<i_l,\, i_s \in \{1,\ldots,n\}\}
$$
for the space $\bigwedge V$ of antisymmetric tensors. We can extend the
order on $V$ to the \emph{inverse lexicographic order} on $\bigwedge V$.
We associate to $e_I$ a \emph{degree} $\vert I\vert$ (number of
generators $e_k$ in $e_I$) and a \emph{grade} (or \emph{parity}) as
$\vert I\vert\!\!\mod 2$. The Grassmann algebra is $\mathbb{Z}$-graded
w.r.t. the degree, that is 
$\bigwedge^i V \bigwedge^j V \subseteq \bigwedge^{i+j} V$. This can be
restricted to a $\mathbb{Z}_2$-grading
$\bigwedge V = \bigwedge_0 V \oplus \bigwedge_1 V$ w.r.t. the grade,
hence decomposing into even and odd parts ($\bigwedge_0 V$ is a
subalgebra, $\bigwedge_1 V$ is a $\bigwedge_0 V$-module). The subspaces
$\bigwedge^i V$ have $\binom{n}{i}$ many basis elements and hence
$\bigwedge V$ has dimension $2^n=\sum_i \binom{n}{i}$. Homogeneous
elements $v_1\wedge \cdots \wedge v_i$ of $\bigwedge^i V$ are called
\emph{extensors} (or \emph{blades}). A general element
$X\in \bigwedge V$ is an aggregate $X = \sum x_I e_I$ with real
coefficients $x_I$. A presentation using generators and relations is
given as 
$$
\bigwedge V:=\langle e_k \mid e_ke_i=-e_ie_k,\,i,k\in \{1,\ldots,n\}\rangle.
$$  

The Grassmann algebra comes with two main involutions. The
\emph{grade involution} extends the map $- : V \rightarrow V :: v\mapsto -v$ 
(additive inverse in $V$) to $(\hat{~})$ on $\bigwedge V$. On generators
this reads as $\hat{e}_I = (-1)^{\vert I\vert}e_I$ and it extends by
linearity. For example, $\hat{e}_1=-e_1$, $\hat{e}_{1,2,3} := 
\hat{e}_1\wedge \hat{e}_2 \wedge\hat{e}_3 = (-1)^3 e_{1,2,3}=-e_{1,2,3}$
while $\hat{e}_{1,2}:= \hat{e}_1\wedge \hat{e}_2 = e_{1,2}$.

A second involution makes use of the opposite algebra. Let $(A,m_A)$
with $m_A : A\times A\rightarrow A :: (a,b) \mapsto ab$ be an algebra.
The \emph{opposite algebra} $A^{op}=(A,m^{op}_A)$ is the same vector
space $A$ with the multiplication
$m^{op}_A : A\times A \rightarrow A :: (a,b) \mapsto ba$. The opposite
wedge product $\wedge^{op}$ is given as $u \wedge^{op} v = v\wedge u$
(no signs), producing the opposite Grassmann algebra $\bigwedge^{op} V$.
The \emph{reversion involution} on $\bigwedge V$ extends the identity
map $1: V\rightarrow V$ to $\bigwedge V$ in such a way that 
$\rev\circ\wedge^{op} = \wedge\circ(\rev\otimes \rev)$. In other words,
$\rev : \bigwedge^{op} V \rightarrow \bigwedge V$ is a Grassmann algebra
isomorphism. As $\rev$ is invertible, this can be used to define
$\wedge^{op}$. On basis elements we have
$\widetilde{e}_I = (-1)^{\vert I\vert(\vert I \vert-1)/2}e_I$ with the
usual notation $\rev(u)=\widetilde{u}$.

Let $V^*=[V,\mathbb{R}]$ be the dual vector space (linear forms). One
finds that $\bigwedge V^* \simeq (\bigwedge V)^*$ is the dual Grassmann
algebra, and we can define a pairing 
$$
\langle-\mid-\rangle : \bigwedge V^* \times \bigwedge V \rightarrow \mathbb{R} :: (e^*_I,e_K) \mapsto \begin{cases} \det(e^*_i(e_k)), & \text{when $|I|=|J|$;}\\
                                     0, & \text{when $|I| \neq |J|$.}
        \end{cases}
$$ 
The \emph{interior product} $\,\JJ\,$ is defined as the adjoint w.r.t.
the duality pairing of multiplication in $\bigwedge V^*$, that is 
$\langle u \mid v\JJ w\rangle := \langle \tilde{v}^* \wedge u \mid w\rangle$.
Let $x,y\in V$ and $u,v,w\in \bigwedge V$ then $\,\JJ\,$ is defined
operationally by the rules (Chevalley construction)
\begin{gather}
\label{eq:interiorproduct}
  x\JJ y =\langle x\mid y\rangle,\;\;
  x\JJ (u\wedge v) = (x\JJ u)\wedge v + \hat{u}\wedge (x\JJ v),\;\;
  (u\wedge v)\JJ w = u \JJ(v \JJ w)\,. 
\end{gather}
Using a nondegenerate bilinear form $B$ as duality, one obtains $x\JJ y = B(x,y)$.

\subsection{(Real) Clifford algebras}\label{subsec:clifford}
Given a quadratic space $(V,Q)$ we can functorially associate to it the
universal Clifford algebra $\CL(V,Q)$ \cite{hm:2008}. Given the
isomorphism $(V,Q)\simeq \mathbb{R}^{p,q}$ this provides the Clifford
algebra $\CL(\mathbb{R}^n,\delta_{p,q})~(n=p+q)$ also often denoted as
$\mathbb{R}^{p,q}$. As vector spaces one has
$\mathbb{R}^{p,q} = \bigwedge\mathbb{R}^n$, hence we have a Grassmann
basis $\{e_I\}$ spanning the vector space underlying the Clifford
algebra. The Clifford product can be implemented in various ways. A
Clifford algebra presentation reads 
$$
\CL_{p,q} :=\langle e_i \mid e_ie_j+e_je_i=0,\, i\not=j;\, e_i^2=1 \textrm{~if~} 1\leq i \leq p, \textrm{~otherwise~} e_i^2=-1\rangle
$$
where the Clifford multiplication is juxtaposition and
$i,j\in\{1,\ldots,n\}$. Another way to define the Clifford multiplication
uses the interior multiplication and the Clifford map 
$\gamma$. Let $x\in V$, $u\in \bigwedge V$ then
$\gamma_x u=xu = x\JJ u + x\wedge u$ and extend by~\eqref{eq:interiorproduct}
with $x\JJ y= B(x,y)$ and linearity. Defining $\CL$ as a quotient of the
tensor algebra it can be seen that it is not graded, but only
$\mathbb{Z}_2$-graded, $\CL=\CL_0\oplus \CL_1$ with $\CL_0$ the even
sub-Clifford algebra and $\CL_1$ a $\CL_0$-module. In terms of the
Grassmann $\mathbb{Z}$-degrees, it is a \textit{filtered algebra}
$\CL^i\CL^j \subseteq \oplus_{r=\vert i-j\vert}^{i+j} \CL^r$. The
quadratic space $(V,Q)$ decomposes as $(V,Q)=(V_1+V_2,Q_1\perp Q_2)$
with restricted quadratic forms $Q_1=Q\vert_{V_1}$ on
$V_1$ and $Q_2=Q\vert_{V_2}$ on $V_2$. The Grassmann algebra functor is
exponential, that is, 
$$
\bigwedge(V_1+V_2) = \bigwedge(V_1)\hotimes\bigwedge(V_2)
$$ 
(the graded tensor product $\hotimes$ is defined below and we use equality
for categorical isomorphisms). Similarly we get for Clifford algebras 
$$
\CL(V_1+V_2,Q_1\perp Q_2)=\CL(V_1,Q_1)\hotimes \CL(V_2,Q_2),
$$ 
and it is this decomposition which will be used below to compute in
\Clifford\ in dimensions  $\ge9$. For Clifford algebras with non-symmetric
bilinear forms such a decomposition is in general \emph{not} direct,
see~\cite{fauser:ablamowicz:2000c}.
\subsection{Tensor products of (graded) algebras}
\label{subsec:tensorproducts}
Let $(A,m_A)$ and $(B,m_B)$ be $\mathbb{K}$-algebras. That is, we have
right and left scalar multiplications $\rho : A\times \mathbb{K}\rightarrow A$
and $\lambda : \mathbb{K}\times B\rightarrow B$. This allows to define
the tensor product $A\otimes_{\mathbb{K}} B$ as a universal object via
a co-equalizer $c$ of two arrows $\rho\circ 1$ and $1\circ\lambda$
$$
  \xymatrix{A \times \mathbb{K} \times B  
  \ar@<0.3ex>[r]^{\rho\circ 1} 
  \ar@<-0.3ex>[r]_{1\circ\lambda} &  A\times B  
  \ar@{->}[r]^{c} &  A\otimes_{\mathbb{K}} B },
$$ 
which produces the common relations such as
$a\lambda\otimes b = a\otimes \lambda b$, multilinearity etc. We have
(vector space) injections $i_A :A \rightarrow A\otimes B$ and $i_B : B
\rightarrow A\otimes B$, and want to transport the algebra structure too.
We define the \emph{product algebra} on $A\otimes B$ from the algebra
structures on $A$ and $B$ as follows: 
$m_{A\otimes B} := (m_A\otimes m_B)(1\otimes \sigma\otimes 1)$, where
$\sigma : A\otimes B\rightarrow B\otimes A :: (a,b) \mapsto (b,a)$ is
the \emph{switch} of tensor factors. On elements:
\begin{gather} 
(a\otimes b) (a^\prime\otimes b^\prime) = (aa^\prime  \otimes bb^\prime)\,. 
\end{gather}
Note that we had to switch $a^\prime$ and $b$, assuming they commute.
As we work in a $\mathbb{Z}_2$-graded setting $A=A_0+A_1$, $B=B_0+B_1$,
this switch has to be replaced with a \emph{graded switch} 
$\hat{\sigma} : A\hotimes B \rightarrow B\hotimes A :: (a,b) 
  \mapsto (-1)^{\vert a\vert\vert b\vert}(b,a)$,
(on homogeneous elements and extended by linearity). The graded
multiplication $m_{A\hotimes B}$ is then given by
$m_{A\hotimes B} := (m_A\hotimes m_B)(1\hotimes \hat{\sigma}\hotimes 1)$.
Using this setup, the injections above become algebra isomorphisms 
$i_A: a \mapsto (a\hotimes 1)$ and $i_B : b \mapsto (1\hotimes b)$. 

In the Grassmann algebra case, splitting the space $V=V_1+V_2$ with $n$
basis vectors $e_i$ into two sets with, respectively, $p$ ($1\leq i\leq p$)
and $q$ ($p<i\leq n$) vectors, we get the maps
$e_i\mapsto e_i\hotimes 1$ ($i\leq p$) and
$e_j\mapsto 1\hotimes e_j$ ($p<j\leq n$). In the CAS computations below
we will \emph{standardize} the indices, that is, we will reindex
$j \mapsto j-p$ so that $i\in\{1,\ldots,p\}$ and $j-p\in\{1,\ldots n-p\}$.
The graded tensor product ensures that we still have the desired
anti-commutation relations
\begin{gather}
  (e_i\hotimes 1)(1\hotimes e_j)
    = (e_i \hotimes e_j)  \quad \mbox{and} \quad (1\hotimes e_j)(e_i\hotimes 1)
    = (-1)^{1\cdot 1} (e_i\hotimes e_j)\,.
\end{gather}
In this way we can form graded tensor products of Clifford algebras
$\CL_{p,q}\hotimes\CL_{r,s}$ too, and that is what we aim for. Tensor
products for matrix algebras are usually called \emph{Kronecker products},
and are taken without the grading. Let
$A\simeq\Mat(n,\mathbb{R})$ and $B\simeq\Mat(m,\mathbb{R})$ be matrix
algebras. Recall that matrices need a choice of basis giving matrices
$[a_{ij}]$ and $[b_{ij}]$. The definition of the matrix tensor algebra
$A\otimes B\simeq\Mat(n\cdot m,\mathbb{R})$ includes a \emph{choice} of
how to form a basis for $A\otimes B$, which consists of elements
$E_{n,m}=e_{i,j}\otimes e_{k,l}$, that is, a reindexing function
$[(i,j),(k,l)] \mapsto (n,m)$. One way is to define a tensor
$[a_{i,j}]\otimes [b_{k,l}] = [ a_{i,j}\cdot[b_{k,l}] ]$ by inserting
the matrix $[b_{k,l}]$ as blocks into the matrix $[a_{i,j}]$; another
obvious choice would exchange the role of $[a]$ and $[b]$. Category
theory shows that the definition of the tensor algebra is unique up to
a unique isomorphism depending on the particular choices. However,
actual computations in a CAS need to be consistent and explicit in the
choice of these isomorphisms.

\subsection{Periodicity of Clifford algebras}
\label{sec:periodicity}
We have seen in the previous section that we can tensor Clifford algebras
of any signature provided we employ the graded tensor product.
\begin{theorem}
Let $\CL_{p,q}$, $\CL_{r,s}$ be two real Clifford algebras, then
\begin{gather}
\CL_{p+r,q+s} \simeq \CL_{p,q}\hotimes \CL_{r,s}
\label{eq:giso}
\end{gather}
(which does not even need nondegeneracy, or even symmetry, of the
involved forms).
\end{theorem}
This result will be used in section~\ref{sec:bigebra} to describe a
general method using \Bigebra\ to do practical symbolic CA computations
in higher dimensional real Clifford algebras of any signature.

Using matrices over Clifford numbers, like $\Mat(2,\CL_{p,q})$, needs
considering \textit{ungraded} tensors, as the matrix algebra tensor
products are ungraded. Doing so employs (graded) algebra isomorphisms
described on the generators of the factor Clifford algebras inside the
ambient Clifford algebra. This leads to the well-known periodicity
relations which are summarized in the following\footnote{%
  For additional references on the periodicity theorems
  see~\cite{angles:2009,atiyah:1964,coquereaux:2009,lam:1973}}.
\begin{theorem}\label{thm:periodicity}
For real Clifford algebras we have the following periodicity theorems
and isomorphisms:
\begin{itemize}
\item[1)] $\CL_{q+1,p-1} \simeq \CL_{p,q}$ if $p \ge 1$ (see \rcite{lounesto:2001}),
\item[2)] $\CL_{p,q+1} \simeq \CL_{q,p+1}$ (see \rcite{port:1995},
\item[3)] $\CL_{p,q}\simeq \CL_{p-4,q+4}$ if $p \ge 4$ (see \rcite{cartan:1908,lounesto:2001}),
\item[4)] $\CL_{p+4,q} \simeq \CL_{p,q}\otimes \CL_{4,0}
           \simeq \CL_{p,q}\otimes \Mat(2,\mathbb{H})$ (see \rcite{port:1995}),
\item[5)] $\CL_{p,q+4} \simeq \CL_{p,q}\otimes \CL_{4,0}
           \simeq \CL_{p,q}\otimes \Mat(2,\mathbb{H})$ (see \rcite{port:1995}),
\item[6)] $\CL_{p+1,q+1} \simeq \CL_{p,q}\otimes \CL_{1,1} \simeq \CL_{p,q}\otimes \Mat(2,\mathbb{R}) \simeq \Mat(2,\CL_{p,q})$ (see \rcite{lounesto:2001}),
\item[7)] $\CL_{p,q+8} \simeq \CL_{p,q}\otimes \Mat(16,\mathbb{R}) \simeq \Mat(16,\CL_{p,q})$ with $\Mat(2,\mathbb{H})\otimes\Mat(2,\mathbb{H})\simeq \Mat(8,\mathbb{R})$  (see \rcite{cartan:1908,lounesto:2001,port:1995}),
\item[8)] $\CL_{p+8,q} \simeq \CL_{p,q}\otimes \Mat(16,\mathbb{R}) \simeq \Mat(16,\CL_{p,q})$ with $\Mat(2,\mathbb{H})\otimes\Mat(2,\mathbb{H})\simeq \Mat(8,\mathbb{R})$  (see \rcite{cartan:1908,lounesto:2001,port:1995}).
\end{itemize}
\end{theorem}
Note that here all tensor products are \emph{ungraded} and these
structure results characterize the cases (signatures) where a
\emph{graded} isomorphism from the graded case to the ungraded case
exists. The isomorphisms with matrix rings need spinor representations
and will be discussed briefly in subsection~\ref{subsec:representations}. 

By way of example we show a typical isomorphism on generators for the
isomorphism 6). 

\begin{example}
Let $\{e_i\}$ be the set of orthonormal generators for $\CL_{p,q}$ with $e_i^2=1$ for 
$i\in\{1,\ldots,p\}$ and $e_i^2=-1$ for $i\in\{p+1,\ldots,p+q\}$, and let $f_1^2=1=-f_2^2$ be the orthogonal generators for $\CL_{1,1}$. Then, 
$e_i \otimes f_1f_2,\, 1\otimes f_1,\, 1\otimes f_2$ form a set of generators for 
$\CL_{p+1,q+1}$.  Indeed, since $f_1f_2=-f_2f_1$ and $(f_1f_2)^2=1,$ for $i,j \in \{1,\ldots, p+q\}$ and $i \neq j$, we find the following familiar relations in $\CL_{p,q}\otimes \CL_{1,1}$:
\begin{gather*}
(e_i \otimes f_1f_2)^2=(e_i \otimes f_1f_2)(e_i \otimes f_1f_2) = e_i^2 \otimes (f_1f_2)^2 = 
\begin{cases} \phantom{-}1 \otimes 1,  & \text{if $1 \leq i \leq p$;}\\
              -1 \otimes 1, & \text{otherwise},
\end{cases}\\
(1 \otimes f_1)^2 = 1 \otimes f_1^2 = 1 \otimes 1 = 
-1 \otimes f_2^2 = -(1 \otimes f_2)^2,\\
(e_i \otimes f_1f_2)(e_j \otimes f_1f_2) + 
(e_j \otimes f_1f_2)(e_i \otimes f_1f_2) = (e_ie_j+e_je_i) \otimes (f_1f_2)^2 = 0,\\
(e_i \otimes f_1f_2)(1 \otimes f_1) + (1 \otimes f_1)(e_i \otimes f_1f_2) = 
e_i \otimes (-f_1^2f_2) + e_i \otimes (f_1^2f_2)=0,\\  
(e_i \otimes f_1f_2)(1 \otimes f_2) + (1 \otimes f_2)(e_i \otimes f_1f_2) = 
e_i \otimes (f_1f_2^2) + e_i \otimes (-f_1f_2^2)=0,\\
(1 \otimes f_1)(1 \otimes f_2)+ (1 \otimes f_2)(1 \otimes f_1) = 
1 \otimes (f_1f_2 + f_2f_1) = 0.    
\end{gather*}
For further details see, e.g.,~\cite{budinich:trautmann:1988a,maks:1989a}.
\end{example}

\begin{theorem}[\cite{budinich:trautmann:1988a}, Thm. 5.8]
\label{thm:bt}
With the notation as above, let $V_2$ have dimension $2k$ and let
$\omega$ be the volume element in $\CL(V_2,Q_2)$ with
$\omega^2=\lambda\not=0$. There exists a vector space isomorphism
between the module $\CL(V_1\oplus V_2,Q_1 \perp Q_2)$ and the module
$\CL(V_1,\frac{1}{\lambda} Q_1)\otimes \CL(V_2,Q_2)$ given on
generators as $(x,y) \mapsto x\otimes \omega + 1 \otimes y$, and there
is a graded algebra isomorphism
\begin{align}
  \CL(V_1\oplus V_2,Q_1 \perp Q_2) 
    \simeq \CL(V_1,\frac{1}{\lambda} Q_1)\otimes \CL(V_2,Q_2).
\label{uiso}
\end{align}
The involutions extend as $(\widehat{x\otimes y}) \simeq \hat{x}\otimes \hat{y}$
and $\rev(x\otimes y) \simeq \rev(x)\otimes\rev(y)$ if $\vert x\vert \equiv 0 \mod 2$
even and $\rev(x\otimes y) \simeq \rev(x)\otimes\rev(\hat{y})$ otherwise.
Then all periodicity isomorphisms in theorem~\ref{thm:periodicity} are
special cases of this one.\footnote{The requirement $\omega^2=\lambda \neq 0$ is equivalent to 
$Q_2$ being non-degenerate. See also~\cite[p. 218]{lounesto:2001}.}
\end{theorem}
To exemplify this, let $(x,y)$ be any pair of generators with
$x \in V_1$ and $y \in V_2$ which upon the embedding 
$V_1 \oplus V_2 \hookrightarrow \CL(V_1 \oplus V_2,Q_1 \perp Q_2)$ we
write as the sum $x+y$. Then,  
\begin{gather}\label{lhs}
  (x+y)^2 = x^2 + (xy+yx) + y^2 
          =(Q_1(x)+Q_2(y))1
          = (Q_1 \perp Q_2)(x,y)
\end{gather} 
due to the orthogonality of $x$ and $y$. On the other hand, in the
(ungraded) tensor product algebra in the right-hand-side of~(\ref{uiso})
we find, as expected,
\begin{align}\label{rhs}
  &(x \otimes \omega + 1 \otimes y)^2 \notag \\% 
  &\hspace*{3ex}= (x \otimes \omega)(x \otimes \omega) + (x \otimes \omega)(1 \otimes y) +
  (1 \otimes y)(x \otimes \omega) + (1 \otimes y)(1 \otimes y)\notag \\
  &\hspace*{3ex}= x^2 \otimes \omega^2 + x \otimes \omega y + x \otimes y\omega + 1 \otimes y^2 \notag \\
  &\hspace*{3ex}= \frac{1}{\lambda}Q_1(x) 1 \otimes \lambda 1 + 1 \otimes Q_2(y)
  = Q_1(x) \otimes 1 + 1 \otimes Q_2(y)\notag \\
  &\hspace*{3ex}= (Q_1(x) + Q_2(y))(1 \otimes 1)
\end{align}
due the anti-commutativity $y \omega = - \omega y$ for every $y \in V_2$
assured by the even dimension of $V_2.$ Furthermore, this last computation
shows that the assumption $\omega^2 = \lambda \neq 0$ and the appearance
of the factor $\frac{1}{\lambda}$ modifying $Q_1$ in $\CL(V_1,\frac{1}{\lambda} Q_1)$,
are both necessary. 

Since we are using Grassmann bases in all Clifford algebras, it is
interesting to calculate the image of a Grassmann basis monomial, lets
say of degree $2$, in the product algebra on the right of~(\ref{uiso}).
Let $x_1,x_2$ be orthogonal generators in $V_1$ and let $y_1,y_2$ be
orthogonal generators in $V_2$. Then, the wedge product in the Clifford
algebra $\CL(V_1\oplus V_2,Q_1 \perp Q_2)$ is computed as expected
\begin{gather}
  (x_1+y_1) \wedge  (x_2+y_2) 
    = x_1 \wedge x_2 + x_1 \wedge y_2 + y_1 \wedge x_2 + y_1 \wedge y_2.
\end{gather}
On the other hand, since $(x_1+y_1) \wedge  (x_2+y_2) = (x_1+y_1)(x_2+y_2)$,
its image in $\CL(V_1,\frac{1}{\lambda} Q_1)\otimes \CL(V_2,Q_2)$ under
the isomorphism~(\ref{uiso}) is the following rather complicated element:
\begin{multline}
  (x_1x_2) \otimes \omega^2 + (x_1 \wedge 1) \otimes \omega y_2 +
  (1 \wedge x_2) \otimes y_1 \omega + (1 \wedge 1) \otimes (y_1 y_2) = \\
  (x_1 \wedge x_2) \otimes \lambda 1 + x_1 \otimes \omega y_2 + 
  x_2 \otimes y_1 \omega + 1 \otimes (y_1 \wedge y_2).
\end{multline}
The isomorphism in~(\ref{uiso}) is given by the procedures
\code{bas2Tbas} (from left to right) and its inverse \code{Tbas2bas}
(from right to left). In the worksheets~\cite{worksheets} we show both
procedures as well as we verify the assertions regarding the involutions.

\subsection{Spinor representations, Clifford valued matrix representations}
\label{subsec:representations}
A Clifford algebra is an abstract algebra, but we may want to realize
it as a concrete matrix algebra. It is, however, well known that matrix
representations may be very inefficient for CAS purposes. The simplest
representation is the (left) regular representation, sending
$a\in A \mapsto \lambda_a = m_A(a,-)\in \End(A)$, the left multiplication
operator by $a$. This representation is usually highly reducible. The
smallest faithful representations of a Clifford algebra are given by
spinor representations. Algebraically, a spinor representation is given
by a \textit{minimal} left ideal which can be generated by left
multiplication from a \textit{primitive} idempotent $f_i=f_i^2$ with
$\not\exists f_k,f_l\not=0$ idempotents such that $f_i=f_k+f_l$ and
$f_kf_l=f_lf_k=0$. The vector space $S:=\CL_{p,q} f_i$ is a
\emph{spinor space}, and it carries a faithful irreducible
representation of $\CL_{p,q}$ for simple algebras.\footnote{%
  For semi-simple Clifford algebras, we realize the spinor representation
  in $S \oplus \hat{S}$ where $\hat{S}=\CL_{p,q}\hat{f}_i$ (see,
  e.g.,~\cite{lounesto:2001}).}
However, when $\CL_{p,q}$ is not simple, and in several signatures
$(p,q)$ this space is not really a vector space, but a module over
$\mathbb{K}=f_i \CL_{p,q} f_i$ with $\mathbb{K}$ isomorphic to
$\mathbb{R}$, $\mathbb{C}$, $\mathbb{H}$, $2\mathbb{R}=\mathbb{R}\oplus\mathbb{R}$,
or $2\mathbb{H}=\mathbb{H}\oplus\mathbb{H}$ depending on the signature
$(p,q)$. The spinor bi-module ${}_{\CL} S_{\mathbb{K}}$ carries a left
$\CL_{p,q}$ and right $\mathbb{K}$ action (scalar product). Looking up
tables of spinor representations~\cite{lounesto:2001, port:1995} yields
that $\CL_{p,q}$ is simple and has a real representation if
$p-q\equiv 0,2 \bmod 8$, distinguished by the fact that a normalized
volume element $\omega$ squares to $\omega^2=+1$ if $p-q \equiv 0,1 \bmod 4$
and $\omega^2=-1$ if $p-q \equiv 2,3 \bmod 4$. Avoiding non real
$\mathbb{K}$'s, we find the isomorphisms 
$\CL_{0,0}\simeq \mathbb{R} \simeq\Mat(1,\mathbb{R})$,
$\{ \CL_{2,0},\CL_{1,1} \} \simeq \Mat(2,\mathbb{R})$,
$\{ \CL_{3,1},\CL_{2,2} \} \simeq \Mat(4,\mathbb{R})$,
$\{ \CL_{4,2},\CL_{3,3},\CL_{0,6} \} \simeq \Mat(8,\mathbb{R})$, and 
$\{ \CL_{8,0},\CL_{5,3},\CL_{4,4},\CL_{1,7},\CL_{0,8}\} \simeq \Mat(16,\mathbb{R})$,
which can be looked up using the command \code{clidata} in \Clifford.
The main reason for using these isomorphisms is that
$\mathbb{R}\otimes \CL_{p,q}\simeq \CL_{p,q}$, so we don't have to
worry about $\mathbb{K}$ tensor products of spinor modules and use only
$S \otimes_{\mathbb{R}} S^* \simeq \Mat(2^k,\mathbb{R})$. 

\begin{example} 
A spinor basis for $\CL_{1,1} \simeq \Mat(2,\mathbb{R})$ used in the
isomorphism 6), given the orthogonal generators $e_1^2=1=-e_2^2$ for
$\CL_{1,1}$ and a primitive idempotent $f$, maybe chosen as   
\begin{gather}
   S=\CL_{1,1}
    =\spn_\mathbb{R} \{f=\frac12(1+e_{1,2}),\, e_1f=\frac12(e_1+e_2)\}.
\end{gather} 
In this basis, the following matrices represent the basis elements in
$\CL_{1,1}$:
%\begin{gather}\label{eq:11spinorrep}
%   [1]   =\left(\begin{array}{cc} 1&0 \\ 0&1 \end{array}\right),\;
%   [e_1] = \left(\begin{array}{cc} 0&1 \\ 1&0\end{array}\right),\;
%   [e_2] = \left(\begin{array}{cc} 0&-1 \\ 1&0 \end{array}\right),\;
%   [e_1e_2] =\left(\begin{array}{cc} 1&0 \\ 0&-1 \end{array}\right)\,.
%\end{gather}
\begin{gather}\label{eq:11spinorrep}
   [1]   = \begin{pmatrix} 1& 0 \\ 0&1 \end{pmatrix},\quad
   [e_1] = \begin{pmatrix} 0&1 \\ 1&0 \end{pmatrix},\quad
   [e_2] = \begin{pmatrix} 0&-1 \\ 1& \phantom{-}0 \end{pmatrix},\quad
   [e_1e_2] = \begin{pmatrix} 1& \phantom{-}0 \\ 0&-1 \end{pmatrix}\,.
\end{gather}
\end{example}

In general, the images of the additional generators $e_i\otimes e_1e_2$
needed to generate $\CL_{p,q}\otimes\CL_{1,1}\simeq \Mat(2,\CL_{p,q})$
are given by $e_i\otimes [e_1e_2]$ with entries from $\CL_{p,q}$, and
read $e_i\otimes e_1e_2 \simeq% 
\begin{pmatrix} e_i & \phantom{-}0 \\ 
0 & -e_i\end{pmatrix}$. 
The isomorphism 6) gives, via iteration, $\CL_{p,q} \simeq \CL_{p-q,0}\otimes \Mat(2^q,\mathbb{R})$ if $p\geq q$ and $\CL_{p,q} \simeq \CL_{0,q-p}\otimes \Mat(2^p,\mathbb{R})$ if $q\geq p$.

\section{Computing with \Clifford\ and \Bigebra\ in tensor algebras}
\label{sec:bigebra}
We provide examples of Maple code how to set up tensor products of
Clifford algebras in \Clifford\ and \Bigebra. For the usage of these
packages see~\cite{A:1996, ablamowicz:fauser:2002d, ablamowicz:fauser:2002c},
the help pages which come with the package, and the package
website~\cite{AF:CLIFFORD}. The Maple worksheets with code for the
described methods will be posted at~\cite{worksheets}.

Loading the package using \code{>with(Clifford);with(Bigebra);} exposes
the exported functions. To set up a Clifford algebra, say $\CL_{2,2}$,
one needs to define the dimension \code{>dim\_V:=2+2;} and the bilinear
form \code{>B:=linalg[diag](1\$2,-1\$2);}.\footnote{%
  The default name of the bilinear form in \Clifford\ and \Bigebra\ is
  $B$, however other names can also be used. So, when the bilinear form
  $B$ is left undefined (unassigned), computations are performed in a
  Clifford algebra $\CL(B)$ for an arbitrary bilinear form $B$ (see,
  e.g.,~\cite{fauser:ablamowicz:2000c,ablamowicz:fauser:2002d}).}
Basis elements $e_I$ are written as strings, e.g., \code{e1we4} stands
for $e_1 \wedge e_4,$ etc., whereas \code{Id} stands for the identity
of the Clifford algebra. \Bigebra\ exports also the (graded) tensor
product \code{\&t}, which is multilinear and associative. Then, a tensor
product $e_{1,2} \otimes e_1$ reads \code{\&t(e1we2,e1)}, and
permutations of tensors are implemented by maps
\code{>switch(\&t(e1,e2),1) = \&t(e2,e1)} (the ungraded switch) or, in
the graded case, \code{>gswitch(\&t(e1,e2),1) = -\&t(e2,e1)} (the graded
switch). The extra index $i$ in either switch (here we have used \code{1}
in each) tells \code{[g]switch} to swap the i-th and the (i+1)st elements.
Again, you get help by typing \code{>?switch} and \code{>?gswitch} at
the Maple prompt.

The Clifford product \code{cmul} by default implicitly uses the bilinear
form $B$ as in, for example, \code{>cmul(e1,e2)=e1we2+B[1,2]*Id}. However,
it can also use~$B$ or any other Maple name explicitly as an optional
argument, e.g., \code{>cmul[K](e1,e2) = e1we2+K[1,2]*Id}, allowing to
compute in different Clifford algebras in the same worksheet. Let
$B$, $B_1$, $B_2$ hold the bilinear forms\footnote{Computations in the worksheet 
\code{cmulGTensor.mw} are performed for \textit{arbitrary} not necessarily symmetric or diagonal bilinear forms.} for $\CL_{p+r,q+s},\CL_{p,q}$
and $\CL_{r,s}$, and let \code{bas2GTbas} be the graded algebra
isomorphism~(\ref{eq:giso}) given by 
$e_I\in\CL_{p,q}\mapsto$\code{\&t(eI,Id)} ($I\subseteq\{1,\ldots,p+q\}$) and 
$e_J\in\CL_{r,s}\mapsto$\code{\&t(Id,eJ)} ($J\subseteq\{1,\ldots r+s\}$),
then the following procedure
\begin{lstlisting}
cmulGTensor:=proc(x,y,B1,B2)
   local f4;
   f4:=(a,b,x,y)->cmul[B1](a,b),cmul[B2](x,y):
   eval(subs(`&t`=f4,gswitch(&t(x,y),2) ));
end proc:
\end{lstlisting}
implements the \emph{graded} tensor product algebra as explained in
section~\ref{subsec:tensorproducts}. We prove in the worksheet \code{cmulGTensor.mw}~\cite{worksheets} that
\newline 
\code{\hspace*{4ex}bas2TGbas(cmulGTensor(X,Y,B1,B2))= \\ 
      \hspace*{40ex}cmul[B](bas2GTBas(X),bas2GTbas(Y))}\\ 
is the graded algebra isomorphism~(\ref{eq:giso}) with the inverse
\code{GTbas2bas}. The grade and reversion involutions work as expected. 
The limit $B\rightarrow 0$ implements the wedge product on $\bigwedge V=\bigwedge V_1\hotimes \bigwedge V_2$.

We discuss in more detail the \emph{ungraded} tensor product for 
$
\CL_{p+1,q+1} \simeq \CL_{p,q}\otimes\CL_{1,1},
$  
the isomorphism 6) of Thm.~\ref{thm:periodicity}, which we call
\code{bas2Tbas} while its inverse is \code{Tbas2bas}. We have in
$\CL_{1,1}$, with generators $e_1^2=1=-e_2^2$, the volume element
$\omega=e_1e_2$ with $\omega^2=\lambda = 1$, as we have to use the
bilinear form $\frac{1}{\lambda} Q_1$ of Thm.~\ref{thm:bt}, so we can
still use $\frac{1}{\lambda} B_1=B_1$. The isomorphism 6) reads
$e_I\in\CL_{p,q}\mapsto$\code{\&t(eI,w)} ($I\subseteq\{1,\ldots,p+q\}$)
and $e_J\in\CL_{1,1}\mapsto$\code{\&t(Id,eJ)}
($J\subseteq\{1,2\}$).\footnote{%
  Here, \code{w} in \code{\&t(eI,w)} stands for the volume element
  $\omega$. Then, in the code of \code{cmulTensor}, the form
  $\frac{1}{\lambda} B_1$ is denoted as \code{lB1}.}
The tensor Clifford product is graded isomorphic to the Clifford product
on $\CL_{p+1,q+1}$. For a proof see the worksheet 
\code{cmulTensor11.mw}.~\cite{worksheets} The following procedure implements  the \emph{ungraded} tensor product algebra as explained in section~\ref{sec:periodicity}.
\begin{lstlisting}
cmulTensor:=proc(x,y,lB1,B2)
   local f4;
   f4:=(a,b,x,y)->cmul[lB1](a,b),cmul[B2](x,y):
   eval(subs(`&t`=f4,switch(&t(x,y),2)));  # ordinary switch
end proc:
\end{lstlisting}
The isomorphism \code{bas2Tbas} is this time more involved, as are
the grade and reversion involutions. We still get the isomorphism 
\newline 
\code{\hspace*{4ex}bas2Tbas(cmulTensor(X,Y,B1,B2))= \\ 
      \hspace*{40ex}cmul[B](bas2TBas(X),bas2Tbas(Y))}\\
proved by direct computation explicitly. Further details are provided
in the worksheet. Note that we do not have the limit
$B\rightarrow 0$, as $Q_2\simeq B_2$ needs to be nondegenerate, and the
naive limit replacing \code{cmul} by \code{wedge} gives false results.
In the worksheet we show how to produce the Grassmann basis for the
tensor algebra, and how the isomorphism and the involutions work.

In each worksheet we have benchmarked computations using the generic \code{cmul} routine
for the Clifford product from \Clifford\ (in $\dim V\leq 9$) versus
the \Bigebra\ tensor routines \code{cmul[G]Tensor}. For orthonormal
bases we roughly get equal run-times. The more complicated data
structures of the tensor algebras is compensated by not computing some
of the off-diagonal terms.

We add to this brief discussion that it is easily possible to iterate
this morphism, and to provide \Clifford\ code for computations in
Clifford algebras 
$$
  \CL_{p+k,q+k}\simeq \CL_{p,q} \otimes \underbrace{\CL_{1,1} \otimes \cdots \otimes \CL_{1,1}}_{k \mbox{ factors}},
$$ 
or use the $\bmod 8$ periodicity.
\section{Computations using matrix algebras over Clifford numbers}
\label{sec:cliffordmatrices} 
The isomorphism 6) from Thm. 2 was explicitly defined by Lounesto 
in~\cite[Sect. 16.3]{lounesto:2001}. We will use this matrix approach to perform computations in 
$\CL_{8,2} \simeq\Mat(2,\CL_{7,1})$~\cite{bayro-corrochano:2012}. Let $\{e_1,\ldots,e_8\}$ be an orthonormal basis of $\mathbb{R}^{7,1}$ generating the Clifford algebra $\CL_{7,1}$ such that 
$e_i^2=1$ for $1\leq i \leq 7$, $e_8^2=-1$, and $e_ie_j = -e_je_i$ for $\leq i,j \leq 8$ and $i \neq j$. The following $2 \times 2$ matrices (compare with~\eqref{eq:11spinorrep}) 
\begin{gather}
E_i = \begin{pmatrix} e_i & \phantom{-}0\\ 0 & -e_i \end{pmatrix} \quad \mbox{for}\quad i=1,\ldots,8, \quad 
E_9 = \begin{pmatrix} 0 & 1\\ 1 & 0 \end{pmatrix}, \quad
E_{10}= \begin{pmatrix} 0 & -1\\ 1 & \phantom{-}0 \end{pmatrix}
\label{eq:repr}
\end{gather}
anti-commute and generate $\CL_{8,2}.$\footnote{Notice that $E_i^2=E_9^2=1$ for $1\leq i \leq 7$ and $E_8^2=E_{10}^2=-1$ in this explicit representation. Also, $\CL_{1,1} \simeq \Mat(2,\mathbb{R})$ is such that the matrices $e_1 = \left(\begin{matrix} 0 & 1\\ 1 & 0 \end{matrix}\right)$ and $e_2=\left(\begin{matrix} 0 & -1\\ 1 & 0 \end{matrix}\right)$ generate $\CL_{1,1}.$} In order to effectively compute in $\CL_{8,2},$ we define the isomorphism $\varphi: \Mat(2,\CL_{7,1}) \rightarrow \CL_{8,2}$ as a Maple procedure \code{phi} from the \code{asvd} package.\footnote{The \code{asvd} package is part of the \Clifford\ library. It was introduced in~\cite{A:2009}.} Its inverse is given by the Maple procedure \code{evalm}. In the first step, we compute 1,024 $2 \times 2$ matrices $E_I$ with entries in $\CL_{7,1}$ which represent the basis monomials $e_I = e_{i_1} e_{i_2} \cdots e_{i_k},$ $i_1 < i_2 < \cdots  < i_k,\, 0\leq k \leq 8$. We store these matrices in a list $\mathcal{B}$. 

For example, for the basis $\mathcal{B}_k$ of $k$-vectors in \Clifford\ we compute all $\binom{8}{k}$ products $E_{i_1}\,\,\code{\&cm}\,\, E_{i_2}\, \code{\&cm}\, \cdots\,\, \code{\&cm}\,\, E_{i_k}$ where \code{\&cm} is a \Clifford\ procedure to compute a product of Clifford algebra-valued matrices with the Clifford product applied to the matrix entries. Once the matrix 
representation~(\ref{eq:repr}) has been chosen,\footnote{Note that other representations are possible. See~\cite[Sect.16.3]{lounesto:2001}.} the list $\mathcal{B}=\bigcup_k \mathcal{B}_k$ can be saved and read into a next Maple session thus avoiding the need to repeat this step. 

Having computed the basis matrices $\mathcal{B}$, we are now ready to compute in $\CL_{8,2}.$ For example,\footnote{In the following, we let $ID$ denote the identity matrix in 
$\Mat(2,\CL_{7,1})$ namely $\begin{pmatrix} \code{Id} & 0\\0 & \code{Id}\end{pmatrix}$ where \code{Id}, as before, denotes the identity element in $\CL_{7,1}.$} let $x = 2ID + 4E_{1,2,3} -10E_{1,5,7,8,10}$ and $y = -ID + 4E_{1,2,3,7} + E_{1,5,6,8} - 3E_{1,4,6,7}$. We can find the Clifford product $x\,\, \code{\&CM}\,\, y$ of $x$ and $y$ in $\CL_{8,2}$ using the following procedure:
\begin{lstlisting}
`&CM`:=proc(x::algebraic,y::algebraic) 
   local xy; global phi,BBB;
   if not type(evalm(x),climatrix) then error 
      `evalm(x) must be of type climatrix` end if:
   if not type(evalm(y),climatrix) then error 
      `evalm(y) must be of type climatrix` end if:
   xy:=displayid(evalm(x) &cm evalm(y));
   return phi(xy,BBB);
end proc:
\end{lstlisting}
Once the basis $\mathcal{B}$ is stored in Maple as a list \code{BBB},
the procedure \code{\&CM} treats it as a global variable and it
returns\footnote{%
On a laptop running Intel(R) Core(TM) 2 Duo CPU T6670 @ 2.20 GHz it
takes 6.5 sec to obtain this result. The computation time can be
shortened using parallel processing available in Maple 15 and later.} 
\begin{multline}
-40E_{2,3,5,8,10}+2E_{1,5,6,8}-16E_{7}-4E_{1,2,3}+10E_{1,5,7,8,10}-10E_{6,7,10}\\
+8E_{1,2,3,7}+4E_{2,3,5,6,8}-12E_{2,3,4,6,7}-30E_{4,5,6,8,10}-6E_{1,4,6,7}-2ID\,.
\end{multline}
Details of the above computations can be found in the worksheet 
\code{G82.mws}~\cite{worksheets}. In the worksheet we also define the respective grade and reversion involutions, and  the graded algebra isomorphisms 
$\CL_{8,2} \leftrightarrow \Mat(2,\CL_{7,1})$ are given by \code{phi} and \code{evalm}.
\providecommand{\bysame}{\leavevmode\hbox to3em{\hrulefill}\thinspace}

\end{document}